\documentclass[11pt,twoside]{article}

\usepackage{asp2006}
\usepackage{epsf}
\usepackage{psfig}
\usepackage{lscape}

\begin{document}
\title{The luminosity function of galaxies in elliptical-dominated galaxy groups: clues on the nature of fossil groups}   
\author{R. Lopes de Oliveira$^1$, C. Mendes de Oliveira$^1$, R. Dupke$^2$, L. Sodr\'e$^1$, E. Cypriano$^1$}   
\affil{1-Universidade de S\~ao Paulo, Brazil; 2-Observat\'orio Nacional, Brazil}    

\begin{abstract} 
We have started a study of luminosity functions of Fossil Group candidates in
order to characterize the faint-end of their galaxy distribution. Here we report
on results of nine of them from SDSS photometry and spectroscopy.
\end{abstract}



Groups of galaxies optically dominated by an elliptical galaxy 
are classified as Fossil Groups (FGs) if the gap in magnitude between the two brightest
galaxies within R$_{virial}$/2 is greater than 2 in the $r$-band,
and L$_{\rm X}$\,$>$\,10$^{42}$ $h_{50}^{-2}$\,erg\,s$^{-1}$. 
We investigated the luminosity functions (LFs) of 9 elliptical-dominated
galaxy groups which are candidates to FGs from SDSS photometry (0.09\,$<$\,$z$\,$<$ 0.15):
J115305.32+675351.5,
J104548.50+042032.5,
J100742.53+380046.6,
J141004.19 +414520.8,
J085640.72+055347.3,
J081526.59+395935.5,
J101745.57+015645.8,
J153950.78+304303.9, and
J171811.93+563956.1
\citep{Koester07,Santos07}.
The main results are:
(i) A Schechter function describes well the individual LFs, with an $\alpha$-parameter ranging from -1.5 to 0.5; two Schechter functions are needed to describe the composite LF, which is characterized by a decrease in number of galaxies around -20\,$<$\,R\,$<$\,-18.5. 
(ii) There is evidence of merging of the brightest galaxies for a given group, given that $\Delta  r$-mag, i.e. the difference in magnitudes between the M* and the mean absolute magnitude computed from the brightest galaxy of all groups is $\sim$ 3 ($\Delta g$-mag\,$\sim$\,2.3 and $\Delta i$-mag\,$\sim$\,3).
(iii) We derive that the faint-end of the composite LF is well fit by $\alpha$\,$>$\,-1.13 in the $r$-band, and $>$\,-0.8 for the $g$ and $i$ bands. These are comparable within errors to parameters derived for the LF of ``normal'' groups. From the available SDSS spectroscopic redshifts and using the \citet{Carlberg97} relation, we estimate 0.7\,$<$\,R$_{virial}$(Mpc)\,$<$\,1.3 for the groups, with a mean value of $\sim$\,1\,Mpc, which corresponds from the SIS model to 4$\times$10$^{13}$\,$<$\,M$_{virial}$(M$_{\odot}$)\,$<$\,2.3$\times$10$^{14}$. 
Future comparisons of the LF of other systems with that of FGs, determined from a larger sample will provide insights into the nature of these systems.

\acknowledgements 
We acknowledge financial support from the Brazilian agencies FAPESP and CNPq (R.L.O.: FAPESP Postdoctoral Research Fellow grant 2007/04710-1). This work is based on observations of the SDSS survey.


\end{document}